# Development of the Listening in Spatialized Noise–Sentences (LiSN-S) Test in Brazilian Portuguese: Presentation Software, Speech Stimuli, and Sentence Equivalence

Bruno MASIERO[(1)], Leticia R. BORGES[(2)], Harvey DILLON[(3)], Maria Francisca COLELLA-SANTOS[(4)]

(1) Communication Acoustics Lab, University of Campinas, Brazil, masiero@unicamp.br
(2) Faculty of Hearing Sciences, Pontifical Catholic University of Campinas, Brazil, leticiarborges@yahoo.com.br
(3) Department of Linguistics, Macquarie University, Australia, harvey.dillon@mq.edu.au
(4) Faculty of Medical Sciences, University of Campinas, Brazil, mfcolell@unicamp.br

**ABSTRACT**
The Listening in Spatialized Noise Sentences (LiSN-S) is a test to evaluate auditory spatial processing currently only available in the English language. It produces a three-dimensional auditory environment under headphones and uses a simple repetition response protocol to determine speech reception thresholds (SRTs) for sentences presented in competing speech under various conditions. In order to develop the LiSN-S test in Brazilian Portuguese, it was necessary to prepare a speech database recorded by professional voice actresses and to devise presentation software. These sentences were presented to 35 adults (aged between 19 and 40 years) and 24 children (aged between 8 and 10 years), all with normal hearing—verified through tone and speech audiometry and tympanometry—and good performance at school. We used a logistic curve describing word error rate versus presentation level, fitted for each sentence, to select a set of 120 sentences for the test. Furthermore, all selected sentences were adjusted in amplitude for equal intelligibility. The framework of LiSN-S in Brazilian Portuguese is ready for normative data analysis. After its conclusion, we believe it will contribute to diagnosing and rehabilitating Brazilian children with complaints related to hearing difficulties in noisy environments.

## 1 INTRODUCTION

Central Auditory Processing (CAP) is defined as the efficiency and effectiveness of the central auditory nervous system in using auditory information. It refers to the perceptual processing of auditory information and the neurobiological activity underlying this processing that gives rise to electrophysiological auditory potentials (1, 2). The analysis and interpretation of auditory information involve several sub-processes or skills necessary for auditory processing to occur and include the neural mechanisms underlying a variety of auditory behaviors, such as sound localization and lateralization; auditory discrimination; auditory pattern recognition; temporal aspects of hearing (integration, discrimination, resolution, and temporal masking); auditory performance on competitive acoustic signals (including dichotic listening) and auditory performance on degraded acoustic signals (3, 4, 5).

Central Auditory Processing Disorder (CAPD) is a dysfunction in the central auditory nervous system (CANS) that leads to hearing difficulties (2). It can occur in patients with neurological damage to the CANS (6), the elderly (7), and children (5). In the pediatric population, there are several possible causes of the disorder, including delayed maturation of the CANS (6), presence of ectopic cells in the auditory system (8), otitis media, as well as CNS injury in a minority of cases (9). The estimated prevalence of this disorder in school-age children ranges from 2 to 5% (4).

Since the 1950s, numerous tests have been created to assess the central functions of hearing. Each test was developed to evaluate a mechanism and a different skill to compose a comprehensive evaluation that allows for correctly quantifying and qualifying the CANS dysfunctions to direct towards a more adequate and efficient rehabilitation.

Several auditory processes involve binaural interactions, including localization in the horizontal plane, and suppression of competing sounds coming from directions different from that of a target sound. This latter process can be called spatial processing (11) and allows us to obtain important information necessary to understand speech and participate in conversations.

The Listening in Spatialized Noise–Sentences (LiSN-S) test (12) was developed in Australia to evaluate

the speech in noise abilities, including the spatial processing abilities of individuals with complaints related to CAPD. It is a speech-in-noise test, applied via dedicated computer software and a headset, producing a three-dimensional virtual auditory environment. The three-dimensional effect is achieved by processing the speech stimulus with the head-related transfer function (HRTF) (12). The test comprises four subtests that use different combinations of spatial arrangement of target and competition, and different combinations of the talkers presenting the target and competing sentences. A simple repetition response protocol is used to obtain the Speech Recognition Threshold (SRT), the signal-to-noise ratio that produces 50% intelligibility for sentences with competitive speech stimuli. LiSN-S is an assessment test sensitive to people with impaired binaural interaction processing (13, 14). Studies have found that the presence of auditory deprivation, such as recurrent episodes of otitis media (OM), may result in decreased ability to use binaural timing and level cues to selectively attend to a target talker in a spatialized listening environment, as assessed by LiSN-S. These studies also highlighted the need for early intervention in these cases and an awareness that the binaural auditory skills may remain in deficit even after sound detection returns to normal. They recommend conducting further studies with children with well-documented otitis media (15, 16).

The primary purpose of this study is to develop the LISN-S test in Brazilian Portuguese, as no test is available to assess spatial processing for native speakers of this language. In this manuscript, we describe the development of a software to assess spatial processing (Section 2), the development of speech material for the Brazilian Portuguese database, which was inserted into the software (Section 3), and the sentence intelligibility equalization procedure (Section 4). The research was approved by the Ethics in Research Committee of Campinas State University under No. 3,462,572, and participants were not reimbursed for their involvement.

## 2  PRESENTATION SOFTWARE

The software and graphical user interface (GUI) were initially developed using the App Designer tool in Matlab. The graphical interface was developed to allow the examiner to enter the patient's data, start the test application procedure, record the number of words correctly repeated by the patient, and visualize the test progress. Based on the data provided by the examiner, the software automatically adjusts the following sentence's sound pressure level to determine the SRT.

Although Matlab's App Designer platform allows the construction of powerful graphical interfaces, the use of Matlab software has two disadvantages: Matlab is a proprietary software with an expensive license and, even running a compiled version of the application, it is necessary to install the Matlab Runtime with over 1 GB in size to run in the background of the main application.

For these reasons, it was decided to rewrite the test application interface in Python[1]. As an interpreted language, Python has a lower computational performance than compiled languages such as C++ and Java. However, this does not become relevant for this specific application because it does not require a large processing power. In this case, the simplicity requirement overrides the performance requirement. For the development of the graphical interface (see an example in Figure 1), the PyQT5 library was chosen, which is compatible with QtDesigner, a software used to design interfaces by dragging and dropping. Audio processing was done with the PyAudio library. Finally, the code was compiled using the cx-Freeze library, which resulted in an executable file of 196 MB, where over half of this file size is composed of audio files.

## 3  AUDIO DATABASE

The sentences used in developing the original LiSN-S were used as a reference for developing the sentences in Brazilian Portuguese. The original sentences were constructed following the criteria for developing the Bamford-Kowal-Bench sentences test (18). The original sentences were translated into Portuguese and had their semantic content, vocabulary, and syntax adapted by a speech pathologist to suit Brazilian children from 6 years onward. The result was 187 sentences, 9 with three words, 49 with four words, 77 with five words, 46 with six words, and 6 with seven words. In addition, two children's stories were selected as competitive speech stimuli.

---

[1] A video illustrating how the LiSN-S test work can be viewed at https://youtu.be/8NBxeIqr8s4.

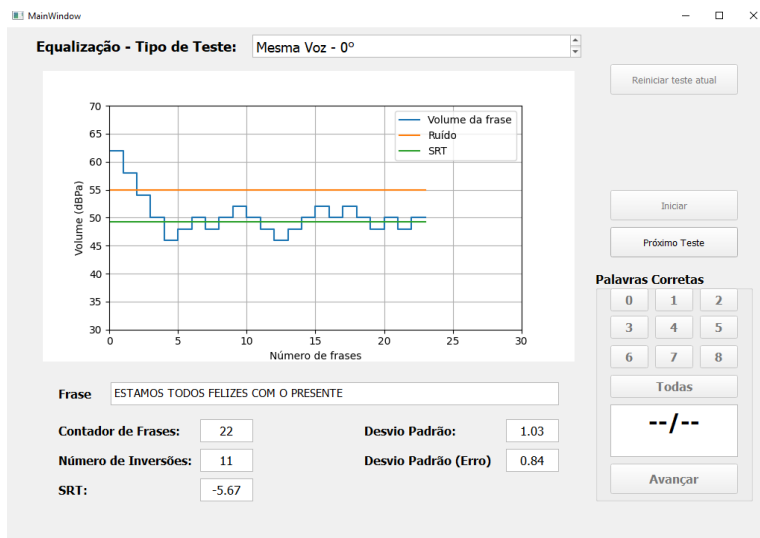

Figure 1. Example of the graphical user interface programmed in Python to administer the LiSN-S tests in Brazilian Portuguese.

The target sentences and competitive stories were recorded in an anechoic chamber by three vocal actresses of the same age group with similar vocal characteristics and accents. They were instructed to speak clearly, with a regular rhythm, and avoid stressing words. Speaker A recorded all target sentences and the two competitive stories, while speakers B and C recorded only the competitive stories. A cardioid microphone, model AKG-c3000, was used for the recording. The recorded material was saved as WAV files using a sampling frequency of 44.1 kHz and 16-bit depth.

The audio material was edited using Audacity. Prolonged pauses were removed during the editing process to ensure the storytelling was smooth and at a constant intensity. Each competitive story lasted just over two minutes, while the target sentences lasted, on average, 2 s. Furthermore, a silent period of 100 ms was inserted immediately before and after each story, while a silent period of 500 ms was inserted immediately before and after each sentence.

The root mean square (RMS) level of all sentences and stories was computed in decibels (dB) and averaged. All signals were then normalized to the calculated average value. After normalization, the stimuli were further attenuated by 7 dB to ensure that saturation did not occur after convolution with the HRTFs. The HRTFs used were measured from a ten-year-old extracted from the openly available Children's HRTFs and Anthropometric Scans for Auditory Research (CHASAR) dataset (17).

## 4 INTELLIGIBILITY EQUALIZATION

The following study was conducted to determine the relative intelligibility of the LiSN-S sentences and to adjust the level of the sentences for equal intelligibility.

### 4.1 Participants

The procedure described above was performed with 35 adults (30 females and 5 males) aged between 19 and 40 years, with a median age of 22 years, and 24 children from a public school (8 female and 16 male) aged between 8 and 10 years, with a median age of 9 years.

Inclusion criteria for control and study groups included:
- normal hearing at the time of assessment, defined as pure tone audiometry thresholds below 20 dBHL in the frequency range of 250 to 8000 Hz;
- normal middle ear function at the time of assessment, measured via tympanometry with peak compliance within 0.3 to 1.3 mʊ at pressure values between -100 to +20 dPa in the presence of 1 kHz ipsi- and contralateral acoustic reflexes in both ears;
- good performance at school, measured by writing and reading tests.

### 4.2 Materials

The hearing tests were performed in a soundproof booth using a two-channel audiometer. The LiSN-S test was applied through a Samsung Windows PC coupled to a Sennheiser HD 280 PRO headset via an RME Madiface audio interface. The headphone presentation level was calibrated using a HEAD Acoustics HSU III.2 head and torso simulator.

### 4.3 Design and Procedure

The equivalence study used the same adaptive procedure used by the LiSN-S test for the "low cue" condition, that is, target sentence and competitive stories presented with the same voice and virtually in front of the listener. Each volunteer was instructed to repeat the phrase heard after the beep and to ignore the competitive story. The number of correctly repeated words was registered into the software.

Target sentences and competing histories were presented simultaneously in six blocks of 31 sentences—to keep the blocks relatively short and thus minimize patient fatigue—with a period of rest in between sections. For the first block, there were an extra two training sentences presented with a fixed signal-to-noise ratio (SNR) of 7 dB. The remaining sentences were presented in an adaptive fashion.

The competitive stimulus was presented at a fixed sound pressure level of 65 dB SPL[21]. Before each target sentence, a warning signal was presented, consisting of a pulsatile tone of 1 kHz and 60 dB SPL, modulated by a Hann window, with a total duration of 200 ms. The sentences were presented 500 ms after the warning. The target sentences were presented in random order. In the first block, there was a training phase consisting of the first three sentences presented at a fixed SNR of 7 dB, i.e., the competitive stimulus is presented at 65 dB SPL, and the target stimulus is presented at 72 dB SPL. From the presentation of the fourth sentence onward, the presentation levels were adjusted in 4 dB steps until the first reversal occurred, that is, the first response with less than 50% of the correct words. If the number of words correctly identified in the previous sentence is exactly 50%, the level of the next target stimulus is kept unaltered. If the first reversal is negative, i.e., the participant responded with less than 50% correct words during training, the block must be restarted. From the first positive reversal onward, sentences were adjusted in 2 dB steps.

At the end of each block, the SRT was calculated as the average of at least three midpoints. In turn, each midpoint was calculated as the average of the level of a positive reversal and its subsequent negative reversal. From the second presentation block onward, the first target sentence was presented at a level 3 dB above the SRT found in the previous block and always using 2 dB steps.

The presentation software recorded the order in which all sentences were presented, the level at which they were presented, and the number of correctly repeated words for each target sentence.

## 5 RESULTS

The data collected, as described in the previous section, was analyzed to obtain a selection method for adequate sentences to compose the Brazilian Portuguese LiSN-S and later correct the selected sentences for equal intelligibility.

### 5.1 Exploratory analysis

For all exploratory statistical analysis that follows, we discarded the first three sentences presented to each participant, as these sentences were presented at a fixed level of 72 dB as a training phase.

We started evaluating the effect of age. A one-way analysis of variance (ANOVA) was performed to compare the effect of age (children VS adults) on the raw target sentence level (TSL). The analysis revealed a significant difference between both groups ($F(1,10854) = 2530.32$, $p = 0$). However, for the intelligibility equalization procedure, we are interested in the adjusted target sentence level (aTSL), i.e., the target sentence level in relation to the individual's estimated SRT. When we repeated the one-way ANOVA to compare the effect of age on the aTSL, the analysis revealed that the difference (0.04 dB) was no longer significant ($F(1,10854) = 0.88$, $p = 0.347$). Therefore, for the remainder of this analysis, we combine both groups.

Next, we analyze the individual performance. To do so, we executed a one-way ANOVA to compare the effect of the individual on the aTSL. The analysis revealed a statistically significant difference in mean aTSL between at least two groups ($F(58,10797) = 1.86$, $p = 8.42e-05$). Tukey's HSD Test for multiple comparisons indicated that two individuals stood out. After discarding the responses of these two individuals, there is no longer any significant difference between the remaining subjects ($F(56,10431) = 1.03$, $p = 0.418$).

Regarding the target sentences, we conducted a one-way ANOVA to evaluate the effect of individual sentences on the aTSL. The analysis revealed a statistically significant difference in mean aTSL between at least two sentences ($F(186,10669) = 3.02$, $p = 0$). Furthermore, a one-way ANOVA on the effect of error rate on the aTSL revealed that the average aTSL differs for the different hit rates ($F(18,10836) = 179.77$, $p = 0$). Both results were expected and corroborated the need for sentence intelligibility equalization. Additionally, we analyzed if the number of words in the target sentence did influence the results. A one-way ANOVA on the effect of the number of words on the aTSL revealed a significant difference between groups ($F(4,10850) = 4.29$, $p = 0.0018$). Tukey's HSD Test for multiple comparisons indicated that the mean value of aTSL was significantly different between sentences with five words and sentences with three or seven words. However, we have no indication that we should rule out a group of sentences by their number of words.

---

[2] Originally, LiSN-S presents the competitive stimulus at 55 dB SPL ([12]). However, as the SRT usually occurs with an SNR of 10 to 15 dB, the threshold target stimulus level is around 40 dB SPL, which can, in itself, make speech recognition difficult.

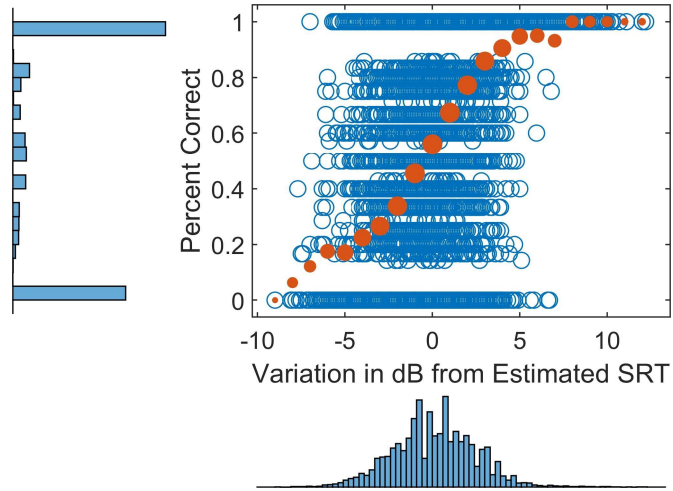

Figure 2. Distribution of aTSL at which the sentence was presented and the rate of correct words. On the left, we have the marginal distribution of the rate of correct words and below the marginal distribution of the aTSL. The orange dots show the average number of correctly recognized words in bands of 1 dB aTSL. The diameter of the points is proportional to the logarithm of the number of presentations.

### 5.2 Distribution of samples

Figure 2 shows the rate of correct words versus the target sentence level relative to the estimated SRT, referred to as aTSL, when combining all 187 target sentences. The histogram at the bottom shows the frequency withwhich sentences were presented at different aTSLs, and it approaches a normal distribution. On the other hand, the histogram on the left appears to have a beta distribution, as it shows that target sentences were often either not recognized (31%) or completely recognized (43%).

For a fixed value of aTSL, the rate of correct words can be modeled as a binomial distribution since the act of correctly recognizing each word can be understood as a Bernoulli experiment, where the output of the experiment is whether the word is correctly recognized or not. The maximum of the binomial distribution occurs at the expected value, given by $E[X] = np$, where $n$ is the number of trials, in our case, the number of words presented at a given aTSL, and $p$ is the probability of correctly recognizing a word at that aTSL.

A logistic curve models the behavior of $p$ as a function of the aTSL.

### 5.3 Sentence Selection

The data collected was used to determine the relative intelligibility of the 187 prepared sentences. Based on that, we were able to select 120 sentences to be used in the Brazilian Portuguese LiSN-S and to adjust the level of these selected sentences for equal intelligibility.

To determine the relative intelligibility, we discarded the training sentences and the responses from two participants, as discussed in Section 5.1. We also discarded three sentences perceived as "bad" during listening tests. For each of the remaining sentences, we combined the results of all volunteers and fit a psychometric function, defined by the logistic curve

$$p = \frac{\exp(a + bX)}{1 + \exp(a + bX)} \qquad (1)$$

where $p$ is the percent correct word and $X$ is the aTSL. One can verify that $S = b/4$ is the slope of the steepest portion of the curve and that the ratio $R = a/b$ represents the SRT, i.e., the aTSL needed to achieve 50% correct identification of words in a sentence. The logistic curves fitted to our data are presented in Figure 3. In contrast, Figure 4 shows the distribution of $S$ and $R$.

Sentences with a "bad" fit, measured by the coefficient of determination $(R^2)$[2] smaller than 0.5, were discarded.

To select the final 120 sentences, we chose those sentences "closest" to the central point of this distribution presented in Figure 4. Among the various possible ways of defining "distance", we chose to use the $\ell_\infty$-norm[3] of the values normalized as a function of their standard deviations. This resulted in the following acceptance region for $S$ and $R$:

- $R-$ median$(R) < 1.9$ dB or $R-$ median$(R) > -1.9$ dB;
- $S < 0.75$ or $S > 0.28$.

---

[2] The coefficient of determination is defined as the proportion of the variation in the dependent variable (in our case, the rate of correct words) that is predictable from the independent variable (in our case the aTSL).

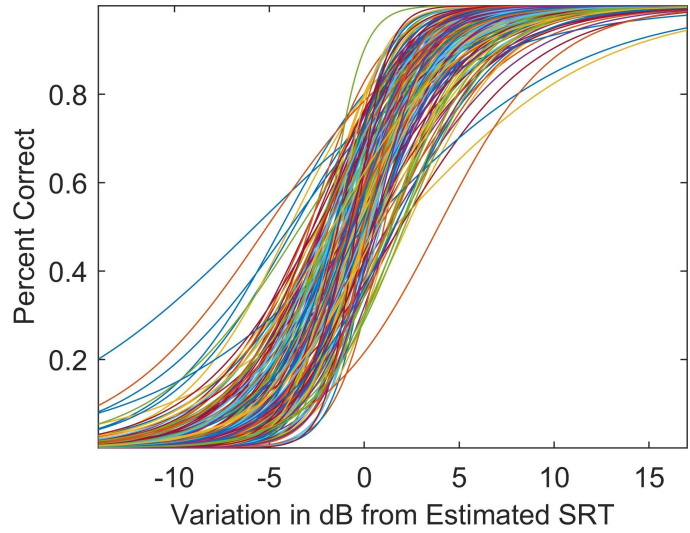

Figure 3. Logistic curves fitted for each of the tested sentences.

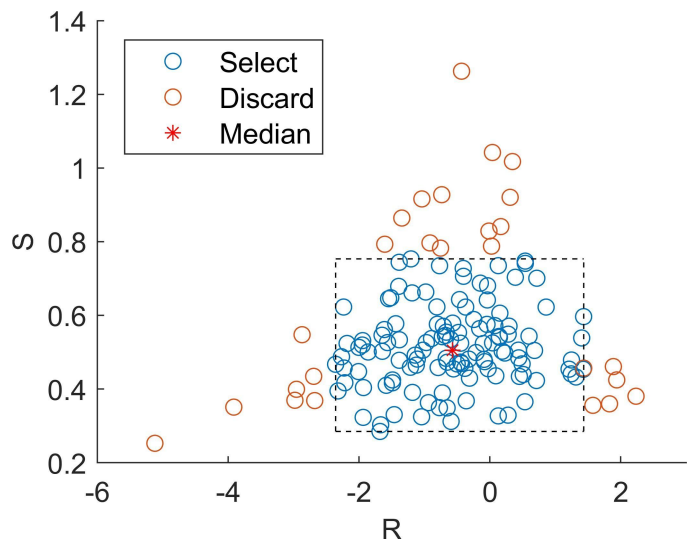

Figure 4. Distribution of R and S values for each of the tested sentences. In red is the median of the values, in blue the points related to the selected phrases, and in orange the discarded sentences.

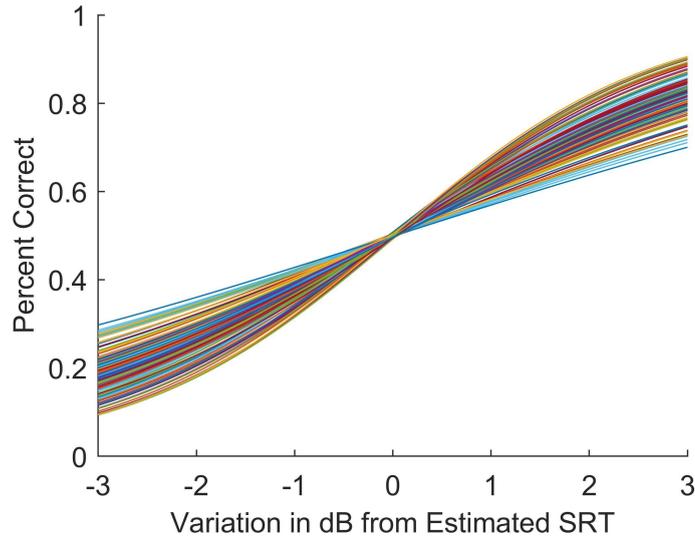

Figure 5. Logistic curves of the 120 selected sentences after adjustment for the same intelligibility.

The selected points are shown in blue in Figure 4, with the discarded ones in orange. Notice how the selection region has a rectangular shape.

### 5.4 Intelligibility adjustment

After selecting 120 sentences, the amplitude of the selected sentences was adjusted to equal intelligibility. To do so, we vary the level of each sentence by the amount needed to make all the psychometric functions have the same level at 50% accuracy. This adjustment is made by amplifying each phrase by $R$, which is equivalent to shifting the curves to the right or left in Figure 3. The result of the curves after adjusting for intelligibility equalization is given in Figure 5.

Equalizing intelligibility, rather than level, should have the effect of increasing the slope of the psychometric function of the test, thus increasing the sensitivity with which the test can reveal abnormalities in speech perception in noise. When the complete set of individual sentence psychometric functions shown in Figure 3 are averaged, the resulting slope at the mid-point of the averaged psychometric function is only 0.11 %/dB. By contrast, after sentence selection and adjustment of level to achieve equal intelligibility, the slope of the function obtained by averaging the curves in figure 5 has increased to 0.13 %/dB.

## 6 CONCLUSIONS

We were able to craft a presentation software for the Brazilian Portuguese version of the LiSN-S test and produce the audio material to be used with the test. After proper audio editing and normalization, sentences were presented to 59 individuals in an adaptive fashion, recording the sentence's presentation level and the number of words repeated correctly. Based on these results, we defined criteria to select 120 sentences as the Brazilian Portuguese LiSN-S sentences and further equalized the selected sentences for equal intelligibility.

Research is ongoing, and software and speech material are ready for application in a selected sample to determine the test's normality criteria. We hope the results of this research will support the effort of understanding the functioning of the central auditory nervous system structures involved in binaural interaction tasks, from the cochlear nucleus to the auditory cortex in Brazilian children. In addition, we hope to disclose the importance of studying spatial processing, especially in children with complaints related to hearing difficulties in noisy environments, to contribute to the diagnosis and help audiologists to plan a more fully and efficient rehabilitation.

---

[3] We could use other distance measures, such as the $\ell_2$ norm, which gives me as a selection region a circle around the median, or the $\ell_1$ norm, whose selection region will be a rhombus, that is, which maximizes the selection of sentences with the same slope and same displacement as the median

## ACKNOWLEDGEMENTS

This work was fully supported by the São Paulo Research Foundation (FAPESP), grants #2017/25267-0 and #2017/08120-6. The opinions, hypotheses, and conclusions or recommendations expressed in this material are the authors' responsibilities, and do not necessarily reflect FAPESP's views.